\documentstyle[emulateapj]{article}
\lefthead{MENOU \& McCLINTOCK} 
\righthead{QUIESCENT EMISSION SPECTRUM OF CEN X-4}

\def\gsim{\;\rlap{\lower 2.5pt
 \hbox{$\sim$}}\raise 1.5pt\hbox{$>$}\;}
\def\lsim{\;\rlap{\lower 2.5pt
   \hbox{$\sim$}}\raise 1.5pt\hbox{$<$}\;} 

\begin{document}

\title{The Quiescent Emission Spectrum of Cen X-4 and other X-ray
Transients containing Neutron Stars} 
\author{Kristen Menou\altaffilmark{1}}

\affil{Princeton University, Department of Astrophysical Sciences,
Princeton NJ 08544, USA, kristen@astro.princeton.edu}

\and

\author{Jeffrey E. McClintock}

\affil{Harvard-Smithsonian Center for Astrophysics, 60 Garden St.,
Cambridge, MA 02138, USA, jem@head-cfa.harvard.edu}

\altaffiltext{1}{Chandra Fellow}

\begin{abstract}
We use the observed optical-UV and X-ray emission spectrum of Cen X-4
during quiescence to constrain models for the accretion flow in this
system. We argue that the optical-UV emission is not due to an
optically-thick quiescent accretion disk, nor due to synchrotron
emission from an Advection-Dominated Accretion Flow (ADAF). Emission
from the bright spot could account for the observed optical-UV
component if the mass transfer rate in Cen X-4 is $\gsim 2 \times
10^{16}$~g~s$^{-1}$. Although the presence of an ADAF around the
neutron star leads to Compton upscattering of the soft X-ray photons
radiated from the stellar surface, we find that this process alone
cannot account for the power law component seen in the quiescent X-ray
spectrum of Cen X-4 and other X-ray transients containing neutron
stars; this result is independent of whether the source of soft
photons is incandescent thermal emission or accretion-powered
emission.  We conclude that, in models which invoke the presence of an
ADAF and a propeller effect for the quiescence of X-ray transients
containing neutron stars, the intrinsic emission from the ADAF must
contribute very little to the optical-UV and X-ray emission
observed. If these ADAF+propeller models are correct, the X-ray power
law component observed must arise from regions where the gas impacts
the neutron star surface. Variability studies could greatly help
clarify the role of the various emission mechanisms involved.
\end{abstract}

\keywords{X-ray: stars -- binaries: close -- accretion, accretion
disks -- stars: neutron -- stars: magnetic fields}

\section{Introduction}

Soft X-ray Transients (SXTs) are close binary systems in which a
low-mass secondary (either a main-sequence star or a subgiant)
transfers mass via Roche-lobe overflow onto a black hole (BH) or
neutron star (NS) primary (see reviews by Tanaka \& Lewin 1995; van
Paradijs \& McClintock 1995; White, Nagase \& Parmar 1995).  SXTs
spend most of their lifetimes in a low luminosity quiescent state, but
occasionally undergo dramatic outbursts during which both the optical
and X-ray emission increase by several orders of magnitude (e.g. Chen,
Shrader \& Livio 1997; Kuulkers 1998). SXTs provide us with some of
the best stellar-mass black hole candidates known to date, thanks to
lower limits set on the mass of the accretor via the observation
during quiescence of Doppler-shifted lines in the spectrum of the
secondary (see, e.g., McClintock 1998).

While observations (see, e.g., Tanaka \& Shibazaki 1996) indicate that
SXTs accrete matter via a standard thin disk (Shakura \& Sunyaev 1973)
during outburst, the situation appears more complex in quiescence
(see, e.g., Lasota 1996).  Narayan, Barret \& McClintock (1997; see
also Narayan, McClintock \& Yi 1996) proposed that accretion in
quiescent BH SXTs proceeds via a two-component accretion flow,
consisting of an inner hot advection-dominated accretion flow (ADAF;
see Narayan, Mahadevan \& Quataert 1998 for a review of ADAFs)
surrounded by an outer thin accretion disk.

According to these models, the unusually low luminosities of BH SXTs
in quiescence is due to the defining property of ADAFs: the bulk of
the viscously dissipated energy is stored in the gas and advected with
the flow into the black hole (Ichimaru 1977; Rees et al. 1982; Narayan
\& Yi 1994, 1995; Abramowicz et al. 1995; Narayan et al. 1996, 1997a).
By contrast, in NS SXTs all the advected energy is expected to be
radiated from the neutron star surface, resulting in a much higher
radiative efficiency of the accretion flow even in the presence of an
ADAF (Narayan \& Yi 1995). This distinction between black hole and
neutron star systems motivated a comparison of the outburst amplitudes
of BH SXTs and NS SXTs as a function of their maximum luminosities by
Narayan, Garcia \& McClintock (1997) and Garcia et al.  (1998). These
authors showed that the observations reveal systematically lower
relative luminosities in BH SXTs and argued that this constitutes a
confirmation of the presence of an event horizon in BH SXTs.

Menou et al. (1999) developed ADAF+thin disk accretion models for
quiescent BH and NS SXTs that account for the luminosity differences
between the two classes of systems (see also Lasota 2000). However,
the relatively low quiescent luminosities of NS SXTs can be reconciled
with the model predictions only if an efficient ``propeller effect''
(Illarionov \& Sunyaev 1975) operates during quiescence, so that the
magnetosphere of the rapidly rotating neutron star prevents much of
the accreting material from reaching the neutron star surface (see
also Asai et al. 1998; Zhang, Yu \& Zhang 1998).

An alternative explanation for the quiescent X-ray emission of NS SXTs
has been proposed by Brown, Bildsten and Rutledge (1998). These
authors argue that nuclear reactions in the crust of neutron stars,
triggered during the outbursts of NS SXTs, efficiently heat up the NS
cores in these systems. The energy deposited during an outburst is
reemitted during quiescence at a rate sufficient to power the observed
quiescent emission (see also Rutledge et al. 1999).  Recently, Campana
\& Stella (2000) also proposed that the relativistic wind of the NS (in
a pulsar regime) in quiescent SXTs could be responsible for part of the
X-ray emission observed; the X-rays are produced by the impact of the
wind on the gas stream flowing from the companion star.

So far, the quiescent emission of NS SXTs has been mainly discussed in
terms of their luminosity, whereas it is likely that spectral
information will help discriminate between the proposed emission
mechanisms discussed above (see, e.g., Rutledge et al. 2000). In this
paper, we construct models for the broadband emission spectrum of
quiescent NS SXTs, with a particular emphasis on Cen X-4. {Our
main goal is to test the ADAF+propeller scenario proposed by Menou et
al. (1999), but some of our results also have implications for the
incandescence scenario put forward by Brown et al. (1998).} An
important motivation for this work is the recent finding by McClintock
\& Remillard (2000) that Cen X-4 (a NS SXT) and A0620-00 (a BH SXT),
despite similar orbital periods, show significantly different
quiescent optical-UV emission. This result could be an indication of
the different nature of the accreting compact object in the two
systems.

In \S2, we recall important observational properties of the NS SXT Cen
X-4, which are later used to constrain our models. In \S3, we discuss
several candidates for the quiescent optical-UV emission of Cen
X-4. We then turn to models of the X-ray spectrum in \S4.  Finally, we
discuss consequences, limitations and possible extensions of this work
in \S5.

\section{Available Constraints}

\subsection{Geometry and Masses}

The orbital period of Cen X-4 is $P_{\rm orb}=15.1$~hr.  The mass of
the secondary is not known precisely but it is clearly small; we take
$M_2=0.15~M_{\odot}$ (White et al. 1995). For the mass of the neutron
star, we adopt $M_1=1.4~M_{\odot}$, corresponding to a binary mass
ratio $q=0.1$ (Charles 1998). The inclination of the system is
$i\simeq 35-40$~deg. (Chevalier et al. 1989; McClinctock \& Remillard
1990).

With these parameters, we estimate an orbital separation $a=2.45
\times 10^{11}$~cm, a circularization radius $R_{\rm circ}=7.5 \times
10^{10}$~cm, and a value $R_{\rm L_1}=1.8 \times 10^{11}$~cm for the
distance from the NS to the $L_1$ Lagrange point (Frank, King \& Raine
1992). Throughout this paper, a distance estimate $d=1.2$~kpc is used
for Cen X-4 (Chevalier et al. 1989; Barret, McClintock \& Grindlay
1996).

According to van Paradijs \& Verbunt (1984), the fluences of the two
known outbursts of Cen X-4 are $3.2 \times 10^{44}$~ergs (July 1969)
and $3 \times 10^{43}$~ergs (May 1979). Using the 1979 outburst
fluence, the 10 years of quiescence and a reasonable value for the
radiative efficiency during outburst ($\eta \equiv L/\dot M c^2=0.1$)
yields an average mass transfer rate $\dot M_T \simeq
10^{15}$~g~s$^{-1}$ for the system. The 10 times larger fluence of the
1969 outburst could indicate a much larger average $\dot M_T$, but we
do not use it here in the absence of a value for the period of
quiescence preceding the outburst.  Note that the mass transfer rate
could also be larger if accretion proceeds via an ADAF at a
significant rate during ($\sim 10$ years of) quiescence but is not
``seen'' because an efficient propeller acts during this phase. The
value $10^{15}$~g~s$^{-1}$ should therefore probably be considered as
a lower limit to $\dot M_T$; in the following sections we consider
values of $\dot M_T \leq 5 \times 10^{16}$~g~s$^{-1}$.

\subsection{Emission spectrum in Quiescence}

An optical-UV spectrum of Cen X-4 has recently been reported by
McClintock \& Remillard (2000). During their observations Cen X-4 was
in quiescence with $V \approx 18.2$; however, on occasion the system
is observed to be as faint as $V \approx 18.6$ (Chevalier et
al. 1989).  The emission is consistent with being roughly flat in $\nu
L_{\nu}$ (with an apparent peak around $\log[\nu({\rm Hz})] \approx
15.1$); it extends up to $\log[\nu({\rm Hz})] \simeq15.2$ (above which
no data are available) for a total luminosity $L_{\rm opt-UV} \simeq
0.8 \times 10^{32}$~ergs~s$^{-1}$.  This is in clear contrast with the
optical-UV spectrum of the BH SXT A0620-00 (of similar $P_{\rm orb}$)
which shows a very strong cutoff at $\log[\nu({\rm Hz})] \simeq 15.1$
(McClintock \& Remillard 2000). The spectra of both systems contain a
strong and relatively broad emission line of Mg~II $2800\AA$.

Two-component, blackbody+power law fits to the quiescent X-ray
emission spectrum of Cen X-4 give a luminosity $L_X \simeq 2-3 \times
10^{32}$~ergs~s$^{-1}$ in the 0.5-10 keV band; $\sim 55$\% of the flux
is in the blackbody (hereafter BB) component of temperature $T_{\rm BB}
\simeq 0.15$~keV, and the rest is in the power law component with
photon index $\alpha_{\rm phot} = 1.9 \pm 0.3$ (Asai et al. 1996; see
also Asai et al. 1998). Similar fits to the quiescent X-ray emission
of the NS SXT Aql X-1 give a luminosity $L_X \simeq 6 \times
10^{32}$~ergs~s$^{-1}$ in the 0.5-10 keV band; $\sim 60$\% of the flux
is in the BB component of temperature $T_{\rm BB} \simeq 0.3$~keV, and the
rest is in the power law component with photon index $\alpha_{\rm
phot} = 1.0 \pm 0.3$ (Campana et al. 1998).

Note that Rutledge et al. (1999) have argued, using fits to neutron
star hydrogen atmosphere models, that the available X-ray spectral
data for NS SXTs in quiescence are consistent with thermal emission
from the entire surface of the neutron star. They find $T_{NS} \simeq
0.1$~keV for Cen X-4 and $T_{NS} \simeq 0.1-0.2$~keV for Aql X-1.

\section{Interpreting the Optical-UV Emission}

We consider three possible candidates for the quiescent optical-UV
emission of Cen X-4 (and other NS SXTs): the emission from a quiescent
disk as predicted by the Disk Instability Model (DIM; see Cannizzo
1993 for a review), the emission from the bright spot, where the
stream of gas from the companion star impacts the outer regions of the
disk, and the ADAF synchrotron emission.

\subsection{Contribution from the quiescent accretion disk}

In the ADAF+thin disk model of Narayan et al. (1997a) for quiescent BH
SXTs (see also Lasota, Narayan \& Yi 1996), only the inner ADAF
contributes to the observed optical and UV emission. In quiescence,
the emission of the truncated disk is primarily in the infrared.  A
similar conclusion applies to NS SXTs if the transition radius between
the inner ADAF and the outer thin disk is also large in these systems
(for BH SXTs, a typical value in the models is $R_{\rm tr}=10^4 R_S$,
where $R_S=3 \times 10^5 (M_1/M_{\odot})$~cm is the Schwarzschild
radius). However, the location of this transition radius is very
uncertain (e.g. Menou et al. 1999); the disk may extend much further
inward.  In this case, could the disk account for or contribute
significantly to the observed optical-UV emission in Cen X-4?

\begin{figure}
\plotone{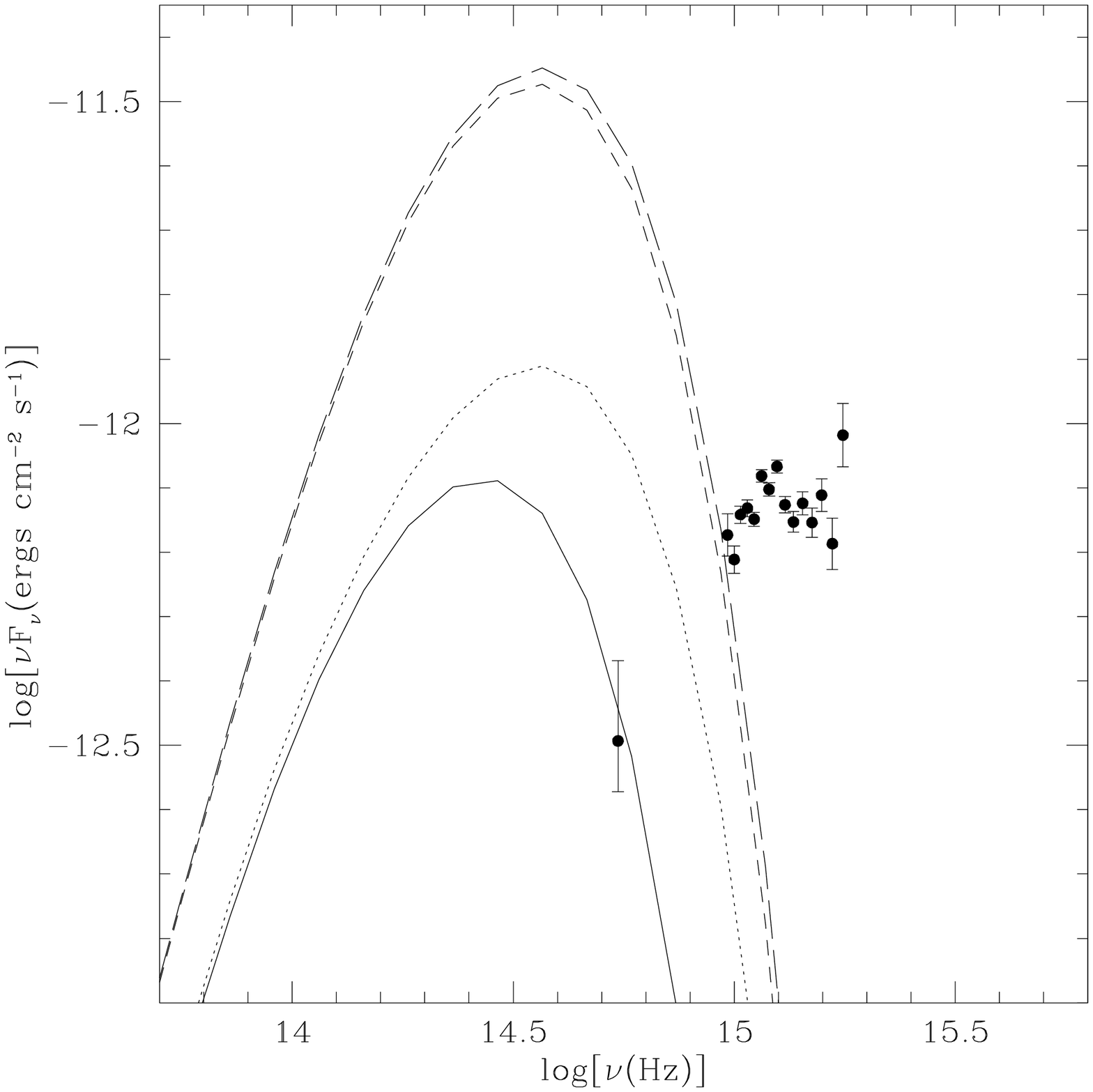}
\caption{Predictions for the emission spectrum of optically-thick
quiescent disks, according to the disk instability model, are compared
to the optical-UV observational data for Cen X-4. The long-dashed line
corresponds to a disk extending from $10^6$ to $10^{11}$ cm, with an
accretion rate $\dot M (R)=min[\dot M_{\rm crit}^-(R),~\dot M_{\rm
out}=5 \times 10^{16}$~g~s$^{-1}]$ (see text for details). The very
similar spectrum (short-dashed line) corresponds to the same disk,
except that the inner radius is $10^8$ cm. The dotted line corresponds
to a disk extending from $10^8$ to $10^{11}$ cm with $\dot M
(R)=min[\dot M_{\rm crit}^-(R),~\dot M_{\rm
out}=10^{16}$~g~s$^{-1}]$. The solid line corresponds to the preceding
model but with $\dot M (R)=min[1/3~\dot M_{\rm crit}^-(R),~\dot M_{\rm
out}=10^{16}$~g~s$^{-1}]$ only. Data points are from McClintock \&
Remillard (2000).
\label{fig:dimdisk}}
\end{figure}

Figure~\ref{fig:dimdisk} shows predictions for the emission spectrum
of a quiescent disk in Cen X-4, according to the DIM. The maximum
accretion rate allowed at any radius in quiescence is given by the
rate at which an annulus would become thermally unstable (see, e.g.,
Hameury et al. 1998):
\begin{equation}
\dot M_{\rm crit}^- \simeq 4 \times 10^{15} \left( \frac{M_1}{M_\odot}
\right)^{-0.89} \left( \frac{R}{10^{10}~{\rm cm}} \right)^{2.67}~{\rm
g~s^{-1}},
\end{equation}
where a very weak dependence on the value of the viscosity parameter
$\alpha$ has been neglected. This accretion rate is essentially the
value at which large opacity changes occur in the disk as the gas
starts becoming ionized.

We assume for the spectral predictions that each annulus is optically
thick and that it emits radiation as a blackbody at the temperature
$T_{\rm eff}$ given by:

\begin{equation}
\sigma T_{\rm eff}^4=\frac{3 G M_1 \dot M}{8 \pi R^3},
\label{eq:teff}
\end{equation}
where boundary condition effects are neglected. Note that the above
relation is strictly valid for a disk in steady-state. At a given
$\dot M$, the emission from a quiescent unsteady disk can be reduced
by up to a factor several compared to that of a steady disk (depending
on the profile of temperature and surface density in the disk; see
Idan et al. 1999). The use of equation~(\ref{eq:teff}) therefore
provides an upper limit to the emission from the quiescent disk. A
reduced emission would only strengthen our conclusions below.

Note that the assumption of large optical thickness is justified for
most of the disk. In the innermost regions, however, this assumption
may not hold. A typical value of the surface density is 
\begin{equation}
\Sigma \lsim \Sigma_{\rm max} \simeq 12~\alpha_{\rm cold}^{-0.83}
(R/10^{10}~{\rm cm})^{1.14}~{\rm g~cm}^{-2},
\end{equation}
where $\alpha_{\rm cold}$ is the viscosity parameter for the quiescent
disk (see, e.g., Hameury et al. 1998). This surface density can be
small enough at small $R$ that the disk is marginally optically thin
in a Rosseland-mean sense.

The four curves in Fig.~\ref{fig:dimdisk} correspond to quiescent disk
models for an inclination $i=40$~deg from the line of sight and a
disk's outer radius of $10^{11}$~cm (see \S2.1).  Given the DIM
constraint on the mass accretion rate, each of the four models is
fully specified by two parameters: the inner disk radius, $R_{\rm
in}$, and the mass transfer rate in the outer part of the disk, $\dot
M_{\rm out}$. {\it Model 1:} $R_{\rm in} = 10^{6} cm$; $\dot M_{\rm
out}= 5 \times 10^{16}$~g~s$^{-1}$. The emission spectrum for this
{\it maximal} disk is indicated by the long-dashed curve in
Fig.~\ref{fig:dimdisk}.  In the inner region, the accretion rate in
the disk is limited to $\dot M(R) = \dot M_{\rm crit}^- (R)$ to
guarantee its stability.  {\it Model 2:} $R_{\rm in} = 10^{8}$~cm;
$\dot M_{\rm out} = 5 \times 10^{16}$~g~s$^{-1}$. The emission
spectrum of this disk model is indicated by the short-dashed curve in
Fig.~\ref{fig:dimdisk}. The disk truncation at $10^8$~cm corresponds
to a reasonable magnetospheric radius in the case of disk accretion at
a rate $\dot M=\dot M_{\rm crit}^- (R)$ onto a neutron star of
magnetic field strength $B=10^8$~G. {\it Model 3:} $R_{\rm in} =
10^{8}$~cm; $\dot M_{\rm out} = 10^{16}$~g~s$^{-1}$. This model is
similar to model~2, except for a reduced value of $\dot M_{\rm out}$
which is consistent with the estimated mass transfer rate in Cen
X-4. {\it Model 4:} $R_{\rm in} = 10^{8}$~cm; $\dot M_{\rm out} =
10^{16}$~g~s$^{-1}$. In this model, indicated by the solid curve in
Fig.~\ref{fig:dimdisk}, the accretion rate in the disk is limited to
$\dot M(R) = 1/3~\dot M_{\rm crit}^- (R)$ in the inner regions. This
is the model which is the closest to the predictions of a full,
time-dependent disk instability calculation.

The models of quiescent disks shown in Fig.~\ref{fig:dimdisk} are
luminous enough to account for the observed quiescent optical-UV
emission of Cen X-4.  However, the spectral discrepancy between the
models and the data is gross.  Even the maximal model is far too faint
in the UV to account for that portion of the spectrum.  This is a more
detailed restatement of the argument made by Wheeler (1996) who
noticed that the blackbody temperatures needed to fit the optical-UV
spectra of quiescent BH SXTs are higher than the maximum values
allowed by the DIM (since such a disk would experience an outburst
immediately).  The argument is stronger for Cen X-4 because the
spectrum extends up to even higher frequencies in the UV.  An
optically-thick disk spectrum that would approximately fit the
optical-UV data in Fig.~\ref{fig:dimdisk} would have an effective
temperature at its inner edge of $T_{\rm in} \sim 20,000$~K, which
would lead to an immediate outburst according to the DIM.  Thus,
assuming that the canonical disk blackbody spectrum is a reasonable
approximation, {we conclude that an optically-thick disk cannot be
responsible for the observed quiescent emission of Cen X-4}.

\subsection{Contribution from the bright spot}

Optical and UV emission from the bright spot, where the stream of gas
transferred by the companion impacts the disk, is seen in dwarf novae
during quiescence; its contribution has been isolated in eclipsing
dwarf novae (Warner 1995).  Since dwarf novae are very similar to
SXTs, except for a primary star which is a white dwarf, some or all of
the quiescent optical-UV emission of Cen X-4 could be due to the
bright spot.

We estimate the maximum luminosity of the bright spot as follows (see,
e.g., Livio 1993).  The gas is essentially free-falling from the $L_1$
point at the surface of the companion star to its impact point on the
disk. Calculations by Lubow (1989) suggest that the minimum radius
reached by such a stream before impact is $\sim 1/2$ the
circularization radius for a low $q$ system such as Cen X-4. Taking
the maximum luminosity of the bright spot to be

\begin{equation}
L_{BS}=G M_1 \dot M_T \left( \frac{1}{0.5 R_{\rm circ}} -
\frac{1}{R_{L_1}} \right),
\end{equation}
we find that the quiescent optical-UV flux of Cen X-4 can be powered
by emission from the bright spot if the mass transfer rate is $\dot
M_T \gsim 2 \times 10^{16}$~g~s$^{-1}~(d/1.2~{\rm kpc})^2$. Since this
is not incompatible with the estimates for $\dot M_T$ discussed in
\S2.1, it seems a plausible explanation for the observed emission. If
the stream impacts the disk at the outer edge, a mass transfer rate a
factor a few times larger would be required. {This possibility,
which is preferred over the two other emission mechanisms considered
here (\S3.1 and \S3.3), could presumably be tested by studying the
variability of the quiescent optical-UV component of Cen X-4.}

We note, however, that the blackbody temperature required to fit the
quiescent optical-UV spectrum of Cen X-4 is $T_{\rm BB} \approx
26,000$~K. This is significantly higher than the values $T_{\rm BB} \simeq
11,000-16,000$~K usually inferred for dwarf novae (Warner
1995). Similarly, the difference between the quiescent optical-UV
spectrum of Cen X-4 and A0620-00 (which would require only $T_{\rm BB}
\simeq 13,000$~K) remains without a clear explanation if the bright
spot hypothesis is adopted.

\subsection{Contribution from the ADAF synchrotron emission}

The optical-UV emission from the accretion flow has been attributed to
the ADAF synchrotron emission in the models for quiescent BH SXTs of
Narayan et al. (1997a).  According to the models of Menou et
al. (1999), an ADAF is present in the inner regions of the accretion
flow of Cen X-4 during quiescence, and mass is mainly flung away at
the magnetosphere by an efficient propeller effect. Some optical-UV
emission is expected from the ADAF (provided it does not possess
strong winds that suppress synchrotron emission; Quataert \& Narayan
1999).  Interestingly, the synchrotron self-absorption cutoff is
expected at higher frequencies for quiescent NS SXTs than for
quiescent BH SXTs because of the smaller mass $M_1$ of the primary
(Mahadevan 1997). The presence of a magnetosphere, however,
complicates the picture, at least by truncating the innermost, hottest
and brightest regions of the synchrotron-emitting accretion flow.

Here, we model the ADAF emission in the simplest way, as it has been
done so far for quiescent BH SXTs (see Narayan et al. 1997a and
Narayan et al. 1998a). We adopt the following model parameters:
$\alpha_{\rm ADAF}=0.1$ (viscosity parameter), $\delta=0.01$ (direct
electron viscous heating), $\beta=10$ (ratio of gas to magnetic
pressure), $p=0$ (no wind) and $i=40$~deg (inclination). The ADAF is
assumed to extend from $R_{\rm out}=10^4 R_S \approx 3 \times 10^9$~cm
down to $R_{\rm in}=10 R_S$ (a reasonable value for the magnetospheric
radius in the case of accretion via an ADAF onto a neutron star of
magnetic field strength $B=10^8$~G, for the typical accretion rates
considered here; see Menou et al. 1999). Since most of the synchrotron
emission comes from the innermost regions of the ADAF, the details of
how the ADAF solution matches to the thin disk are unimportant as long
as the transition radius is large. We neglect here possible
complications due to the ``boundary layer'' at the magnetosphere, and
processes inside the magnetosphere. The soft X-ray emission
originating from the vicinity of the NS (as observationally inferred
for Cen X-4 and other NS SXTs) would only slightly affect the
predicted ADAF emission spectrum in the optical-UV (as shown by
calculations in \S4), so that we neglect it here because it does not
modify our conclusion below.

\begin{figure}
\plotone{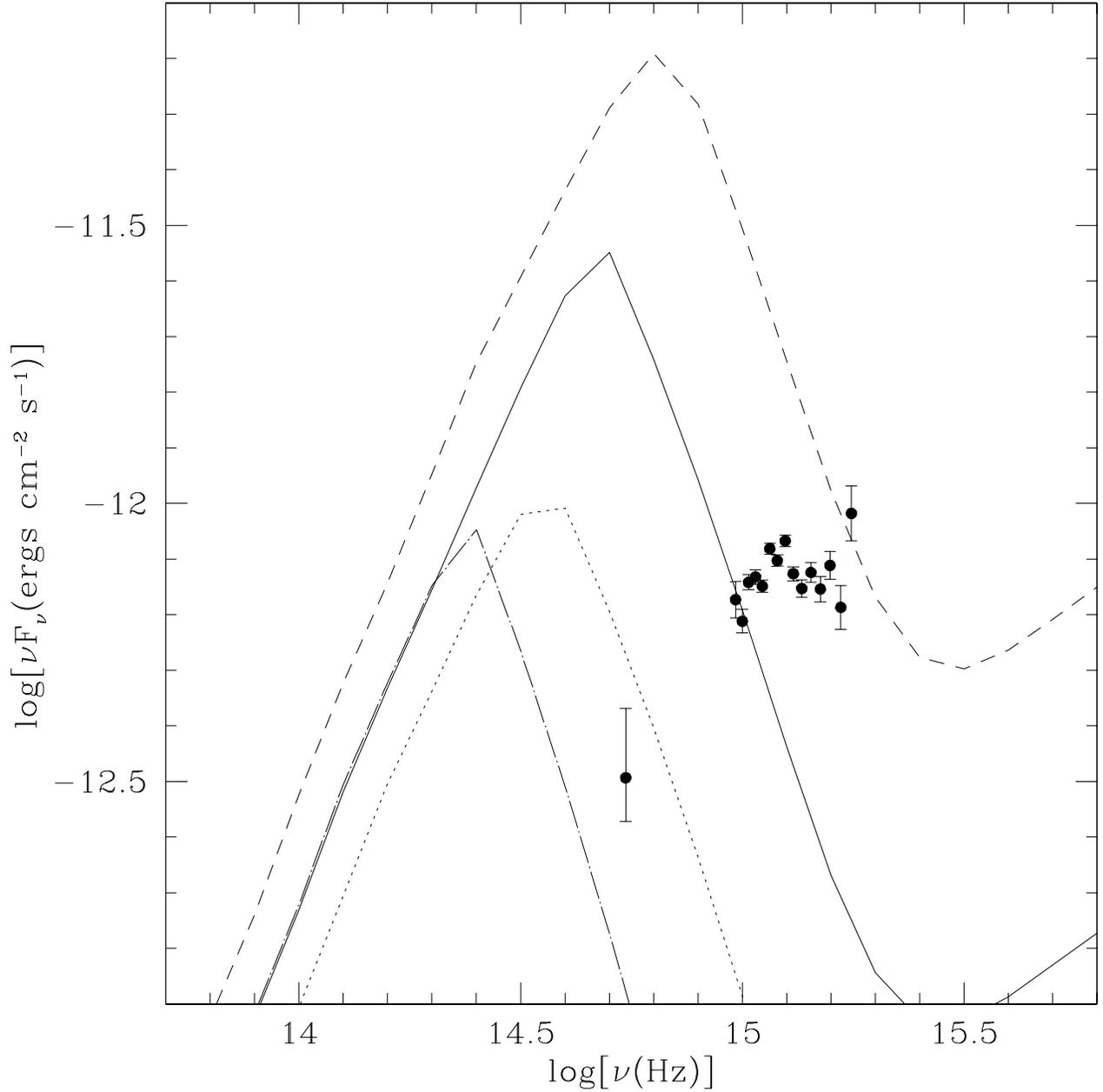}
\caption{Predictions for the synchrotron emission of an ADAF
surrounding the neutron star in Cen X-4 during quiescence. Model
parameters are: $\alpha_{\rm ADAF}=0.1$, $\beta=10$, $\delta=0.01$ and
$p=0$. The three upper curves correspond to models with accretion
rates $\dot m_{\rm ADAF}=2 \times 10^{-3}$ (dashed line; $\approx 4
\times 10^{15}$~g~s$^{-1}$), $10^{-3}$ (solid line; $\approx 2 \times
10^{15}$~g~s$^{-1}$) and $5.5 \times 10^{-4}$ (dotted line; $\approx
10^{15}$~g~s$^{-1}$). The ADAF extends from $10~R_S \approx 3 \times
10^6$~cm to $10^4~R_S$, except for the model represented by the
dashed-dotted line, which has an inner edge at $20~R_S$ (and $\dot
m_{\rm ADAF}=10^{-3}$, as for the solid line). Data points are from
McClintock \& Remillard (2000).
\label{fig:4adafsync}}
\end{figure}

Figure~\ref{fig:4adafsync} shows predictions for the ADAF synchrotron
emission at three accretion rates: $\dot m_{\rm ADAF}= 2 \times
10^{-3}$ (dashed line), $10^{-3}$ (solid line) and $5.5 \times
10^{-4}$ (dotted line) in units of the Eddington accretion rate (taken
here as $\dot M_{\rm Edd}=1.39 \times 10^{18}
(M_1/M_{\odot})$~g~s$^{-1}$); the corresponding physical accretion
rates are approximately $4 \times 10^{15}$~g~s$^{-1}$, $2 \times
10^{15}$~g~s$^{-1}$ and $10^{15}$~g~s$^{-1}$, respectively. None of
the models provides an acceptable description of the data.  The
predicted ADAF synchrotron spectra which are bright enough to account
for the quiescent optical-UV emission are spectrally discrepant, in
the sense that the predicted peak synchrotron frequency is
systematically at a much lower frequency than that at which the
observed spectrum peaks ($\log[\nu({\rm Hz})] \approx 15.1$).

We have checked, by exploring the ADAF parameter space, that this
result is robust to changes in the various model parameters.  The
predicted synchrotron peak frequency is not significantly affected by
variations of $\alpha_{ADAF}$ or $\delta$ (see Narayan et al. 1997a
for an exploration of the parameter space of ADAFs). It is affected by
changes in the parameters $R_{\rm in}$, $\beta$ and $p$, but it turns
out that no model can reproduce at the same time the observed
luminosity and location of the synchrotron peak.  The effect of
varying $R_{\rm in}$ is illustrated by the dashed-dotted line in
Fig.~\ref{fig:4adafsync}, which corresponds to a model with $\dot
m_{\rm ADAF}=10^{-3}$ (as for the solid line) but $R_{\rm
in}=20~R_S$. The synchrotron emission is substantially reduced by the
slight increase of $R_{\rm in}$.  A model with $R_{\rm in}=3~R_S$
(i.e. without magnetospheric truncation of the ADAF), which predicts a
larger peak synchrotron frequency as desired, is ruled out on other
grounds, however. Indeed, without magnetospheric interaction, no
propeller effect is expected, so that the release of all the accretion
energy contained in the hot gas, from the entire NS surface, both
overpredicts by orders of magnitude the observed quiescent bolometric
luminosity of Cen X-4 and dramatically cools the ADAF (strongly
reducing its synchrotron emission) as discussed by Yi et al. (1996).

Reducing the value of $\beta$ (i.e. increasing the magnetic pressure
in the ADAF) leads to larger values of the synchrotron peak frequency,
as shown by Narayan et al. (1997).\footnote{Note the different
definitions of $\beta$ adopted here and in Narayan et al. (1997). For
a given value of $\beta_N$ in Narayan et al. (1997), the corresponding
value of $\beta$ as defined in the present paper is
$\beta=\beta_N/(1-\beta_N)$.} Pushing the value of $\beta$ to unity
(corresponding to an extreme equipartition case) still does not bring
the synchrotron peak to the desired location (see Narayan et al. 1997
for the effect of changing $\beta$). The only model that can account
for the large value of the synchrotron peak frequency is a model with
a necessarily large $\dot m$ (even if $\beta=1$); such a model always
largely overpredicts the observed optical-UV emission of Cen X-4 in
quiescence (see Fig.~\ref{fig:4adafsync} for examples). Even allowing
for the possibility of winds emanating from the ADAF (i.e. models with
$p>0$) does not solve the problem because winds reduce both the
overall synchrotron emission and the value of the synchrotron peak
frequency at the same time (see Quataert \& Narayan 1999). Therefore,
we conclude {ADAF models (that we tested over a wide range of
parameter space) cannot reproduce} the level of optical-UV emission
seen from Cen X-4 in quiescence {\it and} the location of the peak
frequency observed. If there is an ADAF present in the inner regions
of the accretion flow, its emission must be weak enough that it is
swamped by the emission component that produces the observed
optical-UV emission. The ADAF emission can easily be reduced to low
levels if $\dot m$ is low enough and/or the magnetospheric truncation
of the ADAF occurs at large enough radii.

\section{Interpreting the X-ray Emission}

We now consider the quiescent X-ray emission of Cen X-4.  Unlike the
optical-UV emission, we can be quite confident that the X-ray emission
comes from regions near the NS (since coronal emission by the
companion star has been ruled out for quiescent SXTs; Bildsten \&
Rutledge 2000).  There are two competing models for the origin of the
X-ray emission from the neutron star surface: (1) the accretion
luminosity that is powered by the thermal and bulk kinetic energy of
the gas impacting the surface, and (2) the incandescent luminosity due
purely to thermal emission from the intrinsically hot NS (Brown et
al. 1998).
 
The observed power-law component (\S2.2) cannot be due to incandescent
emission, which implies that an accretion flow is responsible for at
least part of the X-ray emission observed.  {In the ADAF+propeller
scenario, a natural origin for this component could be the ADAF itself
(though we argued in \S 3.3 that it cannot be the origin of the
observed quiescent optical-UV emission in Cen X-4).}  Compton
upscattering of the soft X-ray photons by the hot electrons in the
ADAF would produce a hard X-ray component that could be interpreted as
a power-law component in a restricted energy band.

In order to test this hypothesis, we use a modified version of the
ADAF spectral model, developed by R. Narayan, which includes the
possibility of radiation emanating from the NS surface in the spectral
calculations. Apart from this extra feature, the radiative transfer
treatment is identical to that described in Narayan et
al. (1997a). Models are characterized by two extra parameters: $f_{\rm
acc}$ is the fraction of the mass accreted through the ADAF that
reaches the NS surface ($<1$ if a propeller effect is acting in
quiescence) and $f_{\rm surf}$ is the fraction of the NS surface from
which the accretion energy of the gas reaching the NS surface is
radiated to infinity (i.e. $f_{\rm surf}$ essentially characterizes
the geometry of the magnetically channeled accretion onto the NS; the
polar cap axis is assumed to be the same as the axis perpendicular to
the accretion plane).  In this section, we only consider the case of
blackbody-type emission from the NS surface (i.e. we assume that the
energy carried in by the gas has been thermalized before being
radiated away). Given the values of $\dot m_{\rm ADAF}$, $f_{\rm acc}$
and $f_{\rm surf}$, the temperature $T_{\rm BB}$ of the blackbody emission
and the emitted spectrum are easily computed for an assumed NS radius
of 10~km. This emission is assumed to be radiated isotropically from
the NS surface in the Comptonization calculations (see Narayan et
al. 1997a for other details on the radiative transfer treatment).

\begin{figure}
\plotone{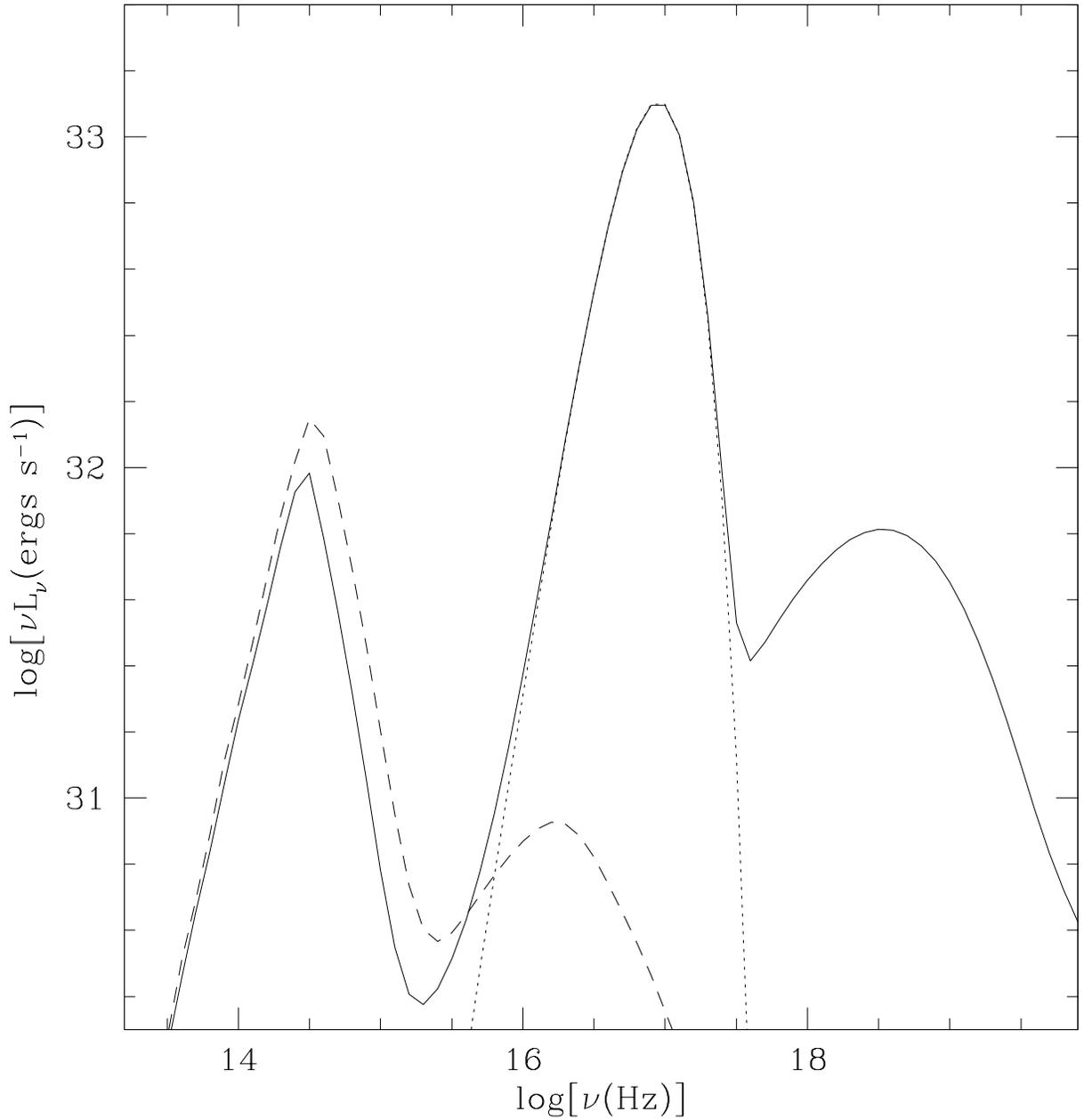}
\caption{Shows the amplitude of Comptonization by the hot
electrons in an ADAF surrounding a neutron star (NS). In this specific
example, the source of soft X-ray photons is blackbody (BB) emission
from the entire NS surface with a temperature $T_{\rm BB}=0.1$ keV (dotted
line). The dashed line corresponds to the original ADAF spectrum, in
the absence of the BB component (model parameters are $\dot m_{\rm
ADAF}=5 \times 10^{-4}$, $\alpha_{\rm ADAF}=0.1$, $\beta=10$,
$\delta=0.01$, $p=0$ and the ADAF extends from $10~R_S \approx 3
\times 10^6$~cm to $10^4~R_S$). The Compton bump which appears when
the BB component is included (solid line) would be interpreted as a
power law component with photon index $\alpha_{\rm phot} \simeq 1.5$
in the 0.5-10 keV energy range. The total energy in the Compton bump
is only a small fraction of that in the BB component because of the
Compton $y$ parameter $\ll 1$ in the ADAF.  
\label{fig:broadspectra}}
\end{figure}

Figure~\ref{fig:broadspectra} shows a representative spectrum at
infinity for a model in which an ADAF surrounds a NS which emits
blackbody-type radiation from its surface (solid line). In this
specific example, the blackbody emission (shown as a dotted line)
emanates from the entire NS surface (i.e. $f_{\rm surf}=1$), with a
temperature $T_{\rm BB}=0.1$~keV. This model represents, at the level of
precision we are interested in here, the emission expected if an
incandescent NS, as proposed by Brown et al. (1998; see Zavlin, Pavlov
\& Shibanov 1996 and Rajagopal \& Romani 1996 for atmosphere models)
were surrounded by an ADAF. There is a clear Compton bump, centered at
$\log[\nu({\rm Hz})]=18.5$, which is due to the Compton upscattering
of the soft X-ray photons of the BB component in the ADAF.  The same
ADAF model without the BB component (dashed line) shows slightly
enhanced synchrotron emission because the cooling of the gas in the
ADAF by Compton upscattering of the soft X-ray photons is absent in
this case.

The Compton bump would be interpreted as a power law component with
$\alpha_{\rm phot} \simeq 1.5$, if it was observed only around
$\log[\nu({\rm Hz})] \sim 18$. An exploration of the parameter space
of the models (i.e. variations of $T_{\rm BB}$ and $f_{\rm surf}$) shows
that exclusively soft ``power law components'', with $\alpha_{\rm
phot} > 1.5$, are produced in models with total X-ray luminosities
comparable to that observed from Cen X-4 in quiescence
($10^{32-33}$~ergs~s$^{-1}$ in the 0.5-10~keV band). Perhaps more
importantly, the fraction of the flux in the Compton component is
invariably small compared to that in the BB component for this type of
models (i.e. $\lsim 5-10$\% bolometric and $\sim$ a few \% in the
0.5-10~keV range, for the model shown in Fig.~3). This is because the
Compton $y$ parameter of the mildly relativistic plasma in the
low-density ADAF, with an electron-scattering optical depth $\tau
\lsim 0.1$ to infinity, is generically small (Rybicki \& Lightman
1979; Narayan et al. 1998b).

For the same reason, the inadequacy of the ADAF to produce the
observed power law component also applies to the accretion models
proposed by Menou et al. (1999), in which the BB component is powered
by the accretion luminosity of the small fraction of mass that
bypasses the centrifugal barrier at the magnetosphere and reaches a
small fraction of the NS surface (i.e. $f_{\rm acc}$ and $f_{\rm surf}
\ll 1$). Our exploration of the parameter space of this class of
models shows that those models which reproduce the typical X-ray
luminosities of NS SXTs in quiescence ($10^{32-33}$~ergs~s$^{-1}$ in
the $0.5-10$~keV band) invariably predict a minor fraction (typically
a few percent) of the flux in the Compton component (hence in the
``power law component''), and values for the photon power law index
$\alpha_{\rm phot} > 1.5$. Given the harder spectrum (smaller
$\alpha_{\rm phot} = 1.0 \pm 0.3$; \S2.2) inferred for Aql X-1 and,
more importantly, the much larger fraction of energy observed in the
power law component of Cen X-4 and Aql X-1 (comparable to the flux
observed in the BB component; \S2.2), we conclude that models which
attribute the X-ray power law component to Compton upscattering in an
ADAF are observationally ruled out. This conclusion is independent of
the origin of the BB component (incandescent thermal emission or
accretion-powered emission).

\section{Discussion and Conclusion}

We constructed models of the accretion flow in the NS SXT Cen X-4
during quiescence. Many of our conclusions can in principle be
generalized to other NS SXTs. We do not find a satisfying explanation
for the observed optical-UV and X-ray emission spectrum of Cen X-4
within the framework of ADAF+propeller accretion models.

We argued that the bright spot can power the quiescent optical-UV
emission of Cen X-4 if the mass transfer rate of the system is $\gsim
2 \times 10^{16}$~g~s$^{-1}$. This is larger than the estimates based
on outburst fluences and therefore suggests, if correct, that a fair
amount of ``invisible'' accretion occurs in quiescence, which could be
due to an efficient propeller acting during this phase. Note that we
did not consider the possibility of reprocessing of X-rays by the
accretion disk for powering the optical-UV flux because the X-ray
luminosity, $L_X \simeq 2-3 \times 10^{32}$~ergs~s$^{-1}$, appears
insufficient to explain the strong optical-UV component, with $L_{\rm
opt-UV} \simeq 0.8 \times 10^{32}$~ergs~s$^{-1}$.

When ruling out emission from a quiescent disk, we neglected possible
deviations from blackbody emission in the innermost regions of the
disk. Most of the observed emission actually originates from the outer
regions ($R^2 T_{\rm eff}^4 \propto \dot M^-_{\rm crit}/R \propto
R^{1.67}$; see Eqs.~[1] and~[2]) which are optically thick according
to the Disk Instability Model (DIM). For this reason, we do not expect
possible complications for the emission in the inner regions of the
disk to be sufficient to explain the spectral discrepancy shown in
Fig.~\ref{fig:dimdisk}. Conceivably, the presence of a disk
chromosphere could induce spectral deviations from blackbody-type
emission but more detailed modeling is required to address this
question. Note that the broad Mg~II line in the spectrum of Cen X-4,
with an integrated energy $L_{\rm MgII} \simeq 2.8 \times
10^{30}$~erg~s$^{-1}$ (McClintock \& Remillard 2000), does suggest
emission from an optically-thin disk chromosphere. The single-peaked
shape can be explained by the superposition of two double-peaked
doublet components separated by $7.2 \AA$, thereby supporting a disk
origin for this line.

We argued that Compton upscattering of soft X-ray photons emitted at
the surface of a NS by an ADAF is not sufficient to explain the power
law component inferred in the quiescent X-ray spectrum of Cen X-4 and
other NS SXTs. Instead, most of the flux in the power law component
could be associated with the accretion-powered emission in higher
density regions close to the NS surface (see, e.g., Shapiro \&
Salpeter 1975; Turolla et al. 1994; Popham \& Sunyaev 2000; Medvedev
\& Narayan 2001; we note, however, than none of these models can be
directly applied to the ADAF+propeller scenario because of
discrepancies in the accretion rate or the geometry assumed).

Compton upscattering by a surrounding ADAF could also have applied to
the scenario of Brown et al. (1998) if, for instance, the propeller
effect is efficient enough in quiescence to prevent all the mass
accreted via the ADAF to reach the NS surface, so that the
accretion-powered luminosity becomes negligible compared to the NS
incandescent emission. However, the difficulties with this scenario
are similar to the case of accretion-powered emission: Compton
upscattering in the low-density ADAF is insufficient to produce the
observed X-ray power law emission. Therefore, the origin of the power
law component in the incandescence scenario remains unclear.

Variability studies could therefore play an important role in
clarifying which mechanisms operate in quiescent NS SXTs. For
instance, variability in the BB component would rule out the scenario
of Brown et al. (1998). Similarly, if coherent X-ray pulsations were
to be observed in a quiescent NS SXT, they would occur in both the BB
and the power law component according to the scenario presented in
\S5, while they should not occur for the BB component according to the
scenario of Brown et al. (1998). Campana \& Stella (2000) have
discussed additional variability characteristics expected in their
pulsar scenario.

It is somewhat embarrassing that no obvious signature of accretion via
an ADAF exists in the model proposed for the quiescent emission
spectrum of Cen X-4. The ADAF remains unseen in optical-UV, perhaps
because of the truncation by the magnetosphere; similarly, its X-ray
emission must be dominated by accretion-powered emission at the
surface of the NS, which has a much higher radiative efficiency than
the ADAF. Consequently, studying the effect of possible winds from the
ADAF (which would only reduce the intrinsic ADAF emission; Blandford
\& Begelman 1998; Quataert \& Narayan 1999) is essentially irrelevant
in the present context. This also means that it is impossible, if
there is indeed an ADAF and a propeller acting in Cen X-4 during
quiescence, to constrain the efficiency of the propeller relative to
that of the winds at removing mass from the accretion flow.

Finally, we note that our study does not offer a clear explanation for
the qualitative spectral difference between A0620-00 and Cen X-4 in
the optical-UV. Although the presence of a magnetosphere in the case
of Cen X-4 is expected to reduce the synchrotron ADAF emission
relative to A0620, and therefore lead to spectral differences, this
does not explain the significantly hotter UV emission of Cen X-4.

\section*{Acknowledgments}

The authors are grateful to Didier Barret, Mike Garcia, Jean-Pierre
Lasota, Ramesh Narayan and Eliot Quataert for useful discussions and
thank R. Narayan for providing the code that he developed for studying
ADAFs around NSs.  Support for this work was provided by NASA through
Chandra Postdoctoral Fellowship grant number PF9-10006 awarded by the
Chandra X-ray Center, which is operated by the Smithsonian
Astrophysical Observatory for NASA under contract NAS8-39073. Partial
support for JM was provided by the Smithsonian Institution Scholarly
Studies Program.

\clearpage

\clearpage

\clearpage

\end{document}